**Title:** Estimation of Multi-Component Flow in the Kidney with Multi-b-value Spectral Diffusion

**Authors:** Mira M. Liu PhD[1], Thomas Gladytz PhD[2], Jonathan Dyke PhD[3], Ian Bolger[1], Jonas Jasse[4], Sergio Calle[1], Tanner Crews[3], Surya Seshan MD[5], Steven Salvatore MD[5], Isaac Stillman MD[6], Thangamani Muthukumar MD[7], Bachir Taouli MD[1,8], Samira Farouk MD[9], Octavia Bane PhD[1,8,★], and Sara Lewis MD[1,8,★]

**Author Affiliations:**
[1]Icahn School of Medicine at Mount Sinai BioMedical Engineering and Imaging Institute, Department of Diagnostic, Molecular and Interventional Radiology, New York, NY, USA.
[2]Berlin Ultrahigh Field Facility (B.U.F.F.), Max Delbrück Center for Molecular Medicine in the Helmholtz Association, Berlin, Germany.
[3]Department of Radiology/Citigroup Biomedical Imaging Center, Weill Cornell Medicine, New York, NY, USA.
[4]Department of Diagnostic and Interventional Radiology, Medical Faculty and University Hospital Düsseldorf, Heinrich-Heine-University Düsseldorf, Düsseldorf, Germany
[5]Department of Pathology, Weill Cornell Medicine, New York, NY, USA.
[6]Department of Pathology, Icahn School of Medicine at Mount Sinai, Mount Sinai Hospital, New York, NY, USA.
[7]Department of Nephrology and Kidney Transplantation Medicine, Weill Cornell Medicine, New York, NY, USA
[8]Department of Diagnostic, Molecular and Interventional Radiology, Icahn School of Medicine at Mount Sinai, Mount Sinai Hospital, New York, NY, USA.
[9]Department of Transplant Nephrology, Icahn School of Medicine at Mount Sinai, Mount Sinai Hospital, New York, NY, USA.

★ O. Bane and S. Lewis are co-senior authors.

**Corresponding Author:**
Sara Lewis, MD
Professor of Diagnostic, Molecular and Interventional Radiology
BioMedical Engineering and Imaging Institute
Icahn School of Medicine at Mount Sinai
E-mail: Sara.Lewis@mountsinai.org
Tel: (212) 241-0712

**Grant Support and Acknowledgements:**

This project was funded by NIH NIDDK R01DK129888 (PI: Lewis/Bane) and supported by National Institutes of Health (NIH) postdoctoral training grant number TL1TR004420 (Fellow: Liu).

**Word Count:** 4682

**ABSTRACT**

**Purpose:** Examine the theory and potential clinical application of estimated intravoxel 'flow' of separated perfusion, tubular flow, and diffusion from multi-b-value DWI in kidney allografts.

**Methods:** Multi-b-value DWI (9 b-values; 0-800 s/mm$^2$) from a kidney cortex is simulated with anisotropic and non-Gaussian (i.e. anomalous) vascular, tubular, and tissue components and analyzed with a Bayesian biexponential, least-squares triexponential, and spectral diffusion MRI. Comparison and application of biexponential, triexponential, and spectral diffusion $fD$ is demonstrated in a two-center study of 54 kidney allografts patients (21F/33M, 48.8±10.5years) and compared to fibrosis (Banff 2017 interstitial fibrosis and tubular atrophy score 0-6 from clinical biopsies of the renal cortex), impaired kidney function (CKD-EPI 2021 eGFR<45ml/min/1.73m$^2$), and proteinuria.

**Results:** Spectral diffusion $fD$ demonstrated strong correlation to input $fD$ of the simulated anisotropic and anomalous components. It agreed with both three-component diffusion ($y = 1.10x - 0.1$, $R^2 = 0.74$) and two-component diffusion ($y = 1.01 + 0.2, R^2 = 0.88$). $fD$ showed similar or improved agreement and correlation to input than individual parameters, and spectral diffusion showed similar or improved agreement than corresponding bi- and triexponential models. In kidney allografts, $fD$ from spectral diffusion showed that higher allograft fibrosis score had higher $fD_{tissue}$ (f-stat=3.86, p=0.02) and that impaired allograft function showed reduced $fD_{tubule}$ (Mann-Whitney U-test=-2.14, p=0.04). Across diagnostic groups of function and fibrosis, $fD_{vascular}$ negatively correlated with proteinuria ($y = -348x + 1144, p = 0.035\ R^2 = 0.82$).

**Conclusions:** Spectral diffusion MRI with multi-Gaussian $fD$ as a flow proxy separated different anomalous and anisotropic diffusion components of perfusion, tubular flow, and tissue diffusion and may hold clinical value in diffusion MRI of kidney pathophysiology.

**Keywords:** diffusion, kidney, kidney disease, MRI

## 1. INTRODUCTION

Diffusion-weighted magnetic resonance imaging (DWI) is a non-invasive non-contrast measure of water proton motion[1,2]. Unlike IV-contrast or spin-labelling perfusion-weighted methods, DWI calculates the amount of intravoxel motion based on the signal decay curve created by motion-caused dephasing at different diffusion-weighting 'b-values'. When a range of multiple b-values is used, the curve may diverge from a standard mono-exponential into a multi-exponential, suggesting signal contribution from components with different diffusion coefficients, i.e. from separate diffusion regimes. Conventional intravoxel incoherent motion (IVIM) assumes a two-component biexponential with 'pseudo-diffusion' and diffusion as the two diffusion regimes[3]. Pseudo-diffusion is often attributed to fast capillary perfusion and measured with low b-values (<200 s/mm$^2$), while slower diffusion is measured with high b-values (~200-1000 s/mm$^2$) and attributed to microstructure (i.e., tissue cellularity, collagen)[3]. However, organs may have more than two physiologic components[4] with diffusion coefficients in different regimes, and be anisotropic and non-Gaussian, e.g. kidneys contain tissue parenchyma (restricted diffusion), tubules (moderate diffusion), and vasculature (fast diffusion)[4-6]. In these cases, multi-component analysis may be of interest, but fitting to a bi- or triexponential in all voxels enforces a fixed number of components[7-9] despite variation in the kidneys. Further, anomalous diffusion in the kidney has been predominately studied through diffusion kurtosis imaging (DKI)[10,11] rather than its effects on IVIM-DWI.

This work presents a data-driven approach to the theory, simulation, and translational application of a multi-Gaussian flux 'flow proxy' $fD$ from a clinically translational multi b-value IVIM-DWI sequence. By approximation as Gaussian, the flow proxy avoids conventional capillary geometry assumptions from the brain[12] and can be applied to different diffusion regimes in the kidney as the product of diffusion coefficient ($D$), and signal fraction ($f$). To test the effects of this approximation, anisotropic and anomalous DWI signal is simulated from diffusion tensors built with literature values, with added Rician noise to match noise measured in images. $f$ and $D$ are fit from the simulated curves using (1) biexponential, (2) triexponential, and (3) spectral diffusion MRI of the kidney cortex[13]. The $fD$ values of distinct diffusion components are then compared to the input parameters of the simulation. Finally, application of the multi-component $fD$ using the three different fitting methods is demonstrated in a two-center study. of kidney transplants.  Kidney allografts were imaged for tissue, tubular, and vascular diffusion components that would be affected by disease with clinical cortical biopsies and urinalysis for reference.

## 2. THEORY

### 2.1 $fD$ as a Flow Proxy of Isotropic Gaussian Diffusion

While $fD^*$ is a proposed measure of local capillary perfusion using conventional IVIM parameters[12,14], in theory $fD$ [volume/time] could apply to any distinct, isotropic, Gaussian diffusing component that is isotropic and Gaussian. The flow proxy can be estimated from the flux of molecules with a given diffusion coefficient $D\left[\frac{mm^2}{s}\right]$ out of a cubic mm per time, independent of geometric assumptions such as segment length or velocity specific to capillary perfusion[12]. This flux can be approximated by integrating the Gaussian probability over a sphere of a diameter of 1mm, setting it to 0.50, and solving for $\sigma^2(t)$[15].

$$\int_0^{2\pi}\int_0^{\pi}\int_0^{r=0.5mm}\frac{1}{\sqrt{(2\pi\sigma^2)^3}}e^{-\frac{r^2}{2\sigma^2}}r^2\sin\theta dr d\theta d\phi = 0.50 \tag{1}$$

Eq. 1 can be solved for $2\sigma^2$ numerically with the erf approximation to return $2\sigma^2 \approx 0.21mm^2$, which is a function of $D$ with Eq. 2: $\sigma = \sqrt{2D\left[\frac{mm^2}{s}\right]t[s]} = 0.32mm$. Solving for $t$ returns the time for half of the original molecules to have diffused out of a cubic mm, as a function of $D\left[\frac{mm^2}{s}\right]$ (Eq. 3): $t[s] = \frac{(0.32[mm])^2}{2D\left[\frac{mm^2}{s}\right]}$. Le Bihan et al.[12] approximated cerebral blood volume with perfusion fraction $f$ representing the fraction of signal due to blood flow, water content fraction ($f_w$) of the volume, and density as $\rho$ in $\left[\frac{g}{ml}\right]$ as Eq. 4: $Volume\ [ml/g] = \frac{f \times f_w}{\rho}\ [ml/g]$. Combining Eq. 3 and Eq. 4 for volume over time with an assumed kidney tissue density of unity ($\rho = 1\left[\frac{g}{ml}\right]$) returns a flow proxy (Eq. 5).

$$Flow = \frac{volume}{time} = \frac{f \times f_w}{1\left[\frac{g}{ml}\right] \times t[s]} \propto fD \tag{5}$$

This product $fD$ may operate as a local flow proxy for any isotropic diffusion component with a known signal fraction $f$ and diffusion coefficient $D$.

## 2.2 Spectral Diffusion of Multi-b-value DWI to get $f$ and $D$

$f$ and $D$ are commonly calculated by fitting multi-b-value DWI decay curves to fixed models like conventional IVIM (biexponential), apparent diffusion coefficient (mono-exponential), or tri-exponential. In comparison, spectral diffusion can fit a flexible number of distinct diffusion components from a multi-b-value decay curve *without* assuming the number of components or starting values. This is of particular use when the number of diffusion components in a voxel is unknown and variable. In spectral diffusion, the $N$-b-value curve is fit to an unconstrained sum of exponentials with non-negative least squares (NNLS) of $M$ logarithmically spaced $D$ values[13,16]. This outputs a spectrum of the contributions of all $M$ exponential basis vectors. A regularization term is used to smooth in the presence of noise and reduce overfitting as the second term in Eq. 6. In this work, $M$ was set to 300, and the regularization parameter was set to $\lambda = 0.1$ after being found to reduce computation time 208-fold with no loss of fit accuracy compared to generalized cross-validation (Supporting Information S1).

$$\chi_r^2 = \min\left[\sum_{n=1}^{N}\left|\sum_{m=1}^{M} s_m e^{-b_n D_m} - y_n\right|^2 + \lambda\sum_{m=2}^{M-1}\left|s_{m+1} - 2s_m + s_{m-1}\right|^2\right] \quad (6)$$

The resulting spectrum has log-normal peaks (i.e. skew and kurtosis of zero in log-space) that correspond to distinct diffusion components with an example spectrum in Fig.1D of a multi-b-value curve in Fig.1C. For a spectrum with amplitude $s_m$ for each of the M=300 diffusion coefficients, a specific i[th] peak's signal fraction is calculated as the area under the peak $f_i = \frac{\sum s_k}{\sum s_{all}}$ for the subset of indices $k \subseteq (1, M)$ within that i[th] peak. The corresponding diffusion coefficient is calculated as the weighted average coefficient of the peak $D_i = \frac{\sum s_k D_k}{\sum s_k}$. Each distinct peak in the spectrum represents a distinct component with a specific signal fraction and diffusion coefficient as demonstrated by Periquito et al[22]. A proxy for the diffusive flow for each component in that voxel can then be calculated using the product of each peak's $fD$ as shown in Eq.5.

## 2.3 Gaussian Approximation of Anomalous Diffusion

In living biological tissue, molecular motion may deviate from Gaussian diffusion. Instead, different molecules may be within different diffusion regimes and follow anomalous diffusion patterns[17,18]. The effect of the different relations between the diffusion coefficient and time has been represented by Hall et al.[19] with an added variable, $\gamma$, to the standard exponential from the Stejskal-Tanner model (Eq. 7a) as a stretched exponential[20]. For a voxel with signal from $N$ anomalous diffusion components, Eq.7a can be summed to create a multi-component DWI decay curve (Eq. 7b)

$$\frac{S_b}{S_0} = e^{-bD} \rightarrow e^{-(bD_A)^{\gamma}} \quad (7a)$$

$$\frac{S_b}{S_0} = \sum_{i=1}^{N} f_{A,i} e^{-(bD_{A,i})^{\gamma_i}} \quad (7b)$$

The value of $\gamma$ describes the diffusion pattern: $\gamma < 1$ is sub-diffusion, $\gamma = 1$ is Brownian diffusion, and $\gamma > 1$ is super-diffusion. In the kidney, within short time durations (gradient duration $<$ $100ms$) capillary microcirculation may be fast and ballistic ($D \geq 0.05$ mm²/s, $\gamma = 2$) as blood is likely to be traveling at a given velocity along capillary segment lengths in the capillary network[21-24]. In comparison, kidney tubules are wider than capillaries with filtration determined by osmotic pressure, and intracellular motion within complex tissues may be restricted sub-diffusion ($D \leq$ $0.005$ mm²/s, $\gamma < 1$)[25]. $f_A$, $D_A$ and $\gamma$ in Eq. 7b may detect microstructural changes associated with kidney disease[5] but accurate capture of these parameters requires a large range of b-values (0 to >>1000s/mm²) with varying gradient durations, and >>3 directions[26-31]. To avoid this, we examine $f_i D_i$ from a standard multi-b-value sequence (b=[0-800s/mm²], 3-directions) as a Gaussian approximation of $f_{A,i} D_{A,i}$, which is, in turn, approximately $f_{A,i} D_{A,i}^{\gamma} b^{\gamma-1}$ for relevant b-values (Eq. 8).

$$\sum_{i}^{N} f_{A,i} e^{-(bD_{A,i})^{\gamma_i}} = \sum_{i}^{N} f_{A,i} e^{-b(D^{\gamma} b^{\gamma-1})} \approx \sum_{i}^{N} f_i e^{-bD_i} \quad (8a)$$

$$f_{A,i}\left(D_{A,i}^{\gamma} b^{\gamma-1}\right) \approx f_{A,i} D_{A,i} \approx f_i D_i \quad (8b)$$

When Eq. 8. is a reasonable approximation, $f_i D_i$ as a Gaussian estimation can still return a relevant flow proxy of perfusion, tubular flow, and diffusion in the kidney[32] within a reasonable scan time for clinical translation.

## 3. METHODS

### 3.1 Visual Example of Gaussian Approximation with Spectral Diffusion in the Kidney

Clinical allograft biopsies are sampled from the kidney cortex making it the most directly translational region of the kidney. As such, an example cortical kidney voxel with tissue, tubular, and vascular components (Fig.1A) was simulated as three isotropic Gaussian diffusion tensors for tissue parenchyma, convoluted tubules, and vasculature (Fig.1B). These input parameters were chosen from literature healthy kidney values[8,9] and the resulting multi-b-value curve (right-hand side of Eq.8a, Eq.9a) was plotted (Fig.1C, black). A stretched exponential was fit to this curve to return anomalous $f_A$ and $D_A$ (left-hand side of Eq.8a, Eq.9b) that would return a similar curve, and overlaid for comparison (Fig.1C, dashed:blue).

$$f_{vasc}e^{-bD_{vasc}} + f_{tubule}e^{-bD_{tubule}} + f_{tissue}e^{-bD_{tissue}} \quad (9a)$$

$$f_{A,vasc}e^{-(bD_{A,vasc})^{\gamma_{vasc}}} + f_{A,tubule}e^{-(bD_{A,tubule})^{\gamma_{tubule}}} + f_{A,tissue}e^{-(bD_{A,tissue})^{\gamma_{tissue}}} \quad (9b)$$

$f_A$ and $D_A$ in Eq. 9b were fit with $\gamma$ fixed; sub-diffusion ($\gamma = 0.85$), tubular flow ($\gamma = 1$), and vascular perfusion ($\gamma = 2$). Since there is limited published data available on multi-b-value anomalous kidney diffusion at $b \leq 800\frac{s}{mm^2}$, $\gamma \approx 0.85$ was chosen based on $\gamma$ in the brain ($\gamma_{WM} \approx 0.76$; $\gamma_{GM} \approx 0.84$)[19] which is more structurally dense than the kidney. The individual diffusion components are plotted in Fig.1C for direct comparison (Gaussian, solid: grey, green, red; $f_i e^{-bD_i}$; anomalous, dashed: grey, green, red; $f_{A,i}e^{-(bD_i)^{\gamma_i}}$).

Diffusion spectra of Eq. 9a and Eq. 9b were plotted for direct comparison (Fig.1D). To demonstrate Eq. 8b, $f_A D_A^{\gamma} b^{\gamma-1}$, $f_A D_A$, and $fD$ were calculated for all three diffusion components and included in Fig.1D. As anomalous diffusion $f_A D_A^{\gamma} b^{\gamma-1}$ is dependent on $b$, the $b$-value in the center of each respective diffusion regime was used (Fig.1C, demarcated regimes; $b_{tissue} = 500$, $b_{tubule} = 100$, $b_{vasc} = 10$ s/mm$^2$). All code in this work was written and run in Python (Python 3.11.4, Anaconda Inc., 2024) and MATLAB (MATLAB Engine API, 2025).

## 3.2 Simulation of Multi-component Anisotropic and Anomalous Diffusion Signal

Multi-b-value DWI signal of the kidney cortex was simulated using 1000 sets of three anisotropic anomalous diffusion tensors. For every set, three diffusion tensors were created to represent vascular diffusion, tubular diffusion, and tissue diffusion components. Each were given input parameters $D_{trace}$, fractional anisotropy (FA), signal fraction $f$, and $\gamma$. $f$, $D_{trace}$, $FA$, and $\gamma$ were chosen from normal distributions of literature values of vasculature, tubular diffusion, and healthy kidney parenchyma if available, or extrapolation from other organs[13,19,33-35] and are listed in Table 1.

Since FA is non-unique, the eigenvectors and eigenvalues of each diffusion tensor were chosen by generating 500 potential sets for each tensor and selecting the closest to the input FA. The eigenvalues were then scaled to return the input $D_{trace}$. This returned a tensor representing a general ellipsoid with the given $D_{trace}$ and approximate FA. To simulate three-directional DWI of anisotropic diffusion, the tensors were rotated randomly in 3D space, and then measured along the global reference frame and averaged for a $D_{approx}$. The $f$, $D_{approx}$, and $\gamma$ of the three diffusion tensors were then used to simulate 9 b-value, 3-direction, three-component anomalous anisotropic diffusion decay curves following Eq. 9b.

Decay curves were generated for the 1000 sets of 3 tensors with Rician noise added as a non-zero mean Gaussian[36], i.e. for DWI decay curve I(b): variance = $\sigma^2 = 0.02$, mean = $\sqrt{I(b)^2 + \sigma^2}$. This noise was added to match the average signal-to-noise ratio (SNR=50) of b=0 images from kidney MRI images after motion correction, averaging, and denoising (Sect.3.4). The same level of noise was added at all b-values, meaning the SNR decreased at higher b-values as the signal decreased, but still met the suggested consensus minimum SNR of 20-30[37]. The effects of artifacts, motion[38], distortion, cardiac pulsatility[39], and b-value discrepancy were not included in the simulation. All simulations in this work were written and run in Python (Python 3.11.4, Anaconda Inc., 2024) and MATLAB (MATLAB Engine API, 2025)

### 3.3 Multi-component Diffusion Model Fitting and Statistical Analysis

Decay curves were fit to (1) a biexponential with a Bayesian approach[40,41] (2) a triexponential with least-squares and (3) an unconstrained sum of exponentials with NNLS for spectral diffusion (Sect.2.2). The Bayesian statistical approach returned $f$, $D$, and $D^*$ and assumed log-priors for the kidney as D=6.2±1, and D*=3.5±1[42]. Both triexponential and spectral diffusion output $f_{tissue}$, $D_{tissue}$, $f_{tubule}$, $D_{tubule}$, $f_{vasc}$ and $D_{vasc}$ as seen in Eq. 9a. The biexponential $\left(\frac{S_b}{S_0} = f e^{-bD^*} + (1-f)e^{-bD}\right)$ output the conventional $(1-f), D, f, D^*$ as $f_{tissue}, D_{tissue}, f_{vasc}, D_{vasc}$ respectively in this work for consistency. The starting values[bounds] of the least-squares triexponential were $f_{tissue} = 0.7[0,1], D_{tissue} = 0.001[0,0.01]$, $f_{tubule} = 0.2[0,1], D_{tubule} = 0.01[0,0.1]$, and $f_{vasc} = 0.1[0,1], D_{vasc} = 0.1[0,0.5]$.

For all methods, the fractions were normalized to sum to one. Spectral diffusion did not require starting values, but the peaks were sorted based on $D$ value. The largest peak closest to $D = 1.8 \times 10^{-3}$ mm²/s, a tissue diffusion literature value between historical biexponential and triexponential fits[8], was labeled the tissue parenchyma peak. Beyond that, peaks were sorted as 0.8 < tissue < 5 ≤ tubule < 50 ≤ vascular for $D$ in $10^{-3}$ mm²/s. Peaks with $D < 0.8\ 10^{-3}$ mm2/s were excluded as either due to anomalous diffusion or diffusion too slow to capture with $b \leq 800$ s/mm². If there was no spectral peak within a diffusion regime (e.g. no vascular peak with $D \geq 50\ 10^{-3}$ mm²/s) both $f$ and $D$ of that component were set to zero.

Average goodness-of-fit was reported. Individual parameters and output $fD$ were correlated against the input parameters ($f_A, D_{A,trace}, f_A D_{A,trace}$) with linear regression. Agreement was determined by median percent difference $\left(\Delta\% = 200 \times \frac{|input-output|}{input+output}\right)$ rather than mean percent error $\left(200 \times \frac{|input-output|}{input}\right)$ to reduce outsize effects of the small and zero-value input parameters in the denominator. Bias was measured with Bland-Altman analysis. For the biexponential, $fD_{vasc} \equiv fD^*$, $fD_{tubule} = 0$, and $fD_{tissue} \equiv (1-f)D$. The same simulation as described above was also run with two diffusion components, with the tubular component set to zero, to mimic a conventional IVIM model of tissue diffusion and capillary perfusion.

### 3.4 Multi-component Flow in Kidney Allografts

#### 3.4.1 Two-Center Kidney Allograft Study

Advanced kidney DWI was run on 4 control volunteers (1F/3M, 38.5±11.8y) and 54 patients (21F/33M, 48.8±10.5y) with kidney allografts from a prospective, IRB-approved HIPAA-compliant two-center study at the Icahn School of Medicine at Mount Sinai and Weill Cornell Medical Center (5R01DK129888). Subjects were instructed to have restricted food, liquids and caffeine intake beginning 4-6 hours prior to scanning. No intravenous contrast agent was administered in this study. Patient demographics and inclusion/exclusion criteria are in Supporting Information S2. Serum creatinine(mg/dL) and proteinuria(mg/24hr) were collected as urinary biomarkers. eGFR was calculated with CKD-EPI 2021 criteria[43]. To reflect single allograft filtration, eGFR≥ 45 ml/min/1.73m² was considered normal/stable function, and eGFR<45

ml/min/1.73 $m^2$ considered impaired function[44]. Interstitial fibrosis and tubular atrophy (IFTA=ci+ct) scores (range, 0-6) were extracted from the clinical biopsy report, prior to imaging, and scored according to the Banff 2017 classification[45] with IFTA>0 considered 'fibrosis'. Clinical biopsies followed standard sampling of the renal cortex.

### 3.4.2 MRI Protocol and Post-Processing

Patients and volunteers underwent 3T MRI (ISMMS: Skyra, Siemens Healthcare, WCMC: Prisma, Siemens Healthcare) respiratory-gated (by liver-dome tracking, pencil-beam navigator) 2D coronal spoiled gradient EPI DWI. This was using the Siemens Advanced Body Diffusion works-in-progress (research pulse sequence) package IVIM-DWI, WIP-990N-syngo-MR-VE11C (b-values=[0,10,30,50,80,120,200,400,800mm$^2$/s]; TR/TE=1500/58ms, voxel-size=2x2x5mm$^3$, 4-directions, monopolar, interleaved, 16-slices, 3-averages, readout bandwidth=1786 Hz/px, iPAT=2). The denoised and motion corrected trace-weighted 9 b-value DWIs were exported directly from the scanner with 'moco-averages', 'moco bvales', 'moco-3D', 'rescale local bias corruption' and iterative denoising[46] (pg.14, WIP-990N-syngo-MR-VE11C). The sequence was not flow-compensated[23,47] and the run time was 7-15 mins depending on patient respiration. Protocol at both sites was set up by the same individual. Six circular regions-of-interest sampling the cortex on a hilar slice were delineated with two each at the upper, middle, and lower pole by an abdominal radiologist with 14 years of experience on the b=0 map. The Bayesian biexponential fit, least-squares triexponential fit, and spectral diffusion analyses described in Sect.3.3 were applied voxel-wise within the kidney ROIs. For each model, any voxel with $R^2 < 0.70$ was excluded from analysis.

### 3.4.3 Multi-component $fD$ Against Urinalysis Biomarkers

Mean $fD_{vasc}$ and $fD_{tubule}$ from spectral diffusion were correlated against proteinuria and eGFR grouped by control volunteers and four diagnoses: (1) "stable allografts" - normal/stable function and no fibrosis, (2) impaired function and no fibrosis, (3) normal/stable function and fibrosis, (4) impaired function and fibrosis. $fD_{tissue}$ was not directly compared to any urinalysis measure. $fD$ of each component was correlated against eGFR and proteinuria by Spearman's rank correlation, and significance was determined as p<0.05.

#### 3.4.4. Multi-component $fD$ Against Allograft Fibrosis

Potential clinical utility of multi-component $fD$ from all multi-component diffusion models was examined and compared across different pathologies. Allografts were grouped by (1) IFTA score for allografts with normal/stable function and (2) kidney function for allografts without fibrosis. Allografts with both fibrosis and impaired function were excluded to avoid confounding trends. Difference in multi-component flow between dichotomized kidney function groups was analyzed with Mann-Whitney U-test, while difference between fibrosis levels was analyzed with one-way ANOVA. Individual parameter $f$ and $D$ were also reported for fibrosis and difference between fibrosis level analyzed with one-way ANOVA. All significance was determined as p<0.05.

## 4. RESULTS

### 4.1 Visual Example of Gaussian Approximation with Spectral Diffusion in Kidney

In Fig.1D, spectral diffusion $f$, and $D$ of the Gaussian diffusion curve agree well with the Gaussian parameters. In comparison, the anomalous spectrum (blue) demonstrates broadened and shifted spectral peaks as $\gamma$ affects the probability distribution shape of diffusion coefficients[48] (Sect.3.1). The tissue sub-diffusion decay is slower than the Gaussian counterpart (Fig.1C, grey) making $f_{A,tissue}$ higher than Gaussian $f_{tissue}$ to compensate. The vascular super-diffusion decay is faster than the Gaussian counterpart (Fig.1C, red) meaning, $f_{A,vasc}$ is smaller than Gaussian $f$ and $D_A$ is higher than Gaussian $D$ to compensate. However, the approximation in Eq. 8b is shown to hold with $f_A D_A \approx f_i D_i$, and both being close to $f_A D_A^{\gamma} b^{\gamma-1}$ (Fig.1D) for all components without recovering $\gamma$ or quantifying deviance like diffusion kurtosis imaging[49].

### 4.2 Simulation of Multi-component Anisotropic and Anomalous Diffusion Signal

Average goodness-of-fit was comparable between biexponential, triexponential, and spectral fits regardless of the number of components ($R^2$=0.839-0.998). The linear regression correlation plots of output $fD$ vs input $f_A D_A$ of anisotropic anomalous diffusion for all fitting methods are shown in Fig.2.

#### 4.2.1. Comparison Between Individual Parameters $f$ and $D$

The linear regression and average percent difference for the individual parameters $f$ and $D$ from the three fitting methods are shown in Table 2. The Bayesian biexponential fit returned the lowest percent differences $\Delta f_{tissue}$ and $\Delta D_{tissue}$ for three component diffusion. This came at the cost of no tubule component and high error of the vascular component. Spectral diffusion in comparison returned higher $\Delta f_{tissue}$ and $\Delta D_{tissue}$, but better agreement of tubule and vascular components even than triexponential fit. The triexponential returned false non-zero $f_{tubule}$ and $D_{tubule}$ values for the two-component fit, contributing to higher error in the vascular and tissue components.

### 4.2.2 Three-component $fD$ Simulation

Linear regression of output $fD$ to input $f_A D_A$ and corresponding Bland-Altman mean difference, and $\Delta fD$ are shown in Table 3. Spectral diffusion returned the mean difference closest to zero and better correlation and lower $\Delta fD$ than the triexponential. The biexponential demonstrated comparably strong correlation as spectral diffusion, but with a larger mean difference. It returned lower $\Delta fD_{tissue}$ than both spectral and triexponential fit at the cost of no tubule component, demonstrating the trade-off of fitting more than two components (Fig.2, Table 3). With a three-component model, $\Delta fD_{tissue}$ was higher than $\Delta f_{tissue}$ and $\Delta D_{tissue}$ for all three fitting methods. However, linear regression of combined $fD$ (Fig.2, Table 3; $R^2 = 0.67 - 0.74$) compared to individual parameters (Table 2; $R^2 = 0.17 - 0.22$) supports combination improving the correlation.

### 4.2.3. Two-component $fD$ Simulation

For the two-component simulation, the biexponential and spectral diffusion $fD$ outperformed the triexponential that was forced to fit a non-existent component (Fig.2, Table 3). Spectral diffusion had a mean difference closest to zero, slope closest to unity, and the smallest $\Delta fD$ while Bayesian biexponential returned a higher error of the vascular component. The triexponential fit returned the worst agreement due to being forced to fit a false component. However, 266/1000 triexponential fits had two coefficients with $\leq$ 10% mean difference from each other, suggesting the two exponents representing a single component, and 469/1000 with $fD < 0.0001$ suggesting a component that is essentially zero.

### 4.3 Spectral Diffusion Applied in Kidney Allografts

Spectral diffusion parameter maps from one *in vivo* healthy volunteer (M/47y) kidney are shown in Figure 3A-C; these include voxel-wise $f, D$ and $fD$ of each of the three diffusion components.

The DWI decay curve of two example voxels from the *in vivo* kidney is shown with biexponential, triexponential and spectral diffusion fits in Fig.3(D,F). The corresponding spectra are shown in Fig.3(E,G). Like the simulation, the triexponential returned a 'false' tubule component with a diffusion coefficient close to the vascular component in a 'two-component' *in vivo* voxel (Fig.3D).

### 4.3.1 Multi-component $fD$ Against Kidney Biomarkers

Spectral diffusion $fD$ showed significant monotonic correlation with kidney filtration biomarkers averaged across clinical subgroups. $fD_{tubule}$ positively correlated with eGFR in ml/min/1.72m$^2$ ($fD_{tubule} = 0.01eGFR - 0.1, p = 0.005, R^2 = 0.95$) and $fD_{vasc}$ negatively correlated with proteinuria in mg/24hr ($fD_{vasc} = -348Proteinuria + 11.45, p = 0.035, R^2 = 0.82$). Correlation was also observed between $fD_{vasc}$ and eGFR ($fD_{vasc} = -0.01eGFR + 1.9, R^2 = 0.55$), and $fD_{tubule}$ and proteinuria ($fD_{tubule} = 0.03Proteinuria + 4.5, R^2 = 0.75$) although these were not significant (p=0.149, 0.056).

### 4.3.2 Multi-component $fD$Against Allograft Fibrosis

Spectral diffusion multi-component $fD$ showed significant differences for fibrosis scores (Fig.4A-C) and kidney function (Fig.4D-F). Higher fibrosis scores showed significant increase in $fD_{tissue}$ (F-stat=4.18, p=0.015; Fig.4C). Allografts with impaired function and no fibrosis showed decreased $fD_{tubule}$ (Mann-Whitney U-test stat=-2.06, p=0.039; Fig.4E). $fD_{vasc}$ and $fD_{tubule}$ demonstrated decreasing median values at higher fibrosis scores, but they were not significantly different.

Biexponential $fD$ did not return significant difference for different fibrosis levels or kidney function (Fig.5), though it showed trends of increased $fD_{tissue} = (1 - f)D$ with fibrosis and decreased $fD_{vasc} = fD^*$ with impaired function. Triexponential $fD$ showed the higher $fD_{tubule}$ for the highest fibrosis score (Fig.6A-B) which was different from the spectral diffusion (Fig.4A-B), though this may be due to the limited sample size of allografts with both IFTA=6 and normal function. Triexponential $fD_{tissue}$ was not significant against IFTA but showed a trending increase (Fig.6C), and $fD_{tubule}$ was significantly lower in patients with impaired function (Fig.6E).

### 4.3.3. Individual $f$ and $D$ Against Allograft Fibrosis

Individual parameters $f$ and $D$ showed expected trends, but mean values were not significantly different for fibrosis scores (Table 4). In terms of trends, spectral diffusion $f_{vasc}$ showed a consistent decrease (each subsequent IFTA score had a lower mean value) at higher fibrosis scores while $f_{tissue}$ increased. $D_{vasc}$ showed a consistent decrease at higher fibrosis scores, while $D_{tissue}$ showed mixed trends. Biexponential $D^*$ also showed a consistent decrease while $D$ showed a consistent increase. All triexponential parameters showed mixed trends with no consistent increase or decrease across IFTA scores.

## 5. DISCUSSION

This work demonstrated using $fD$ from multi-b-value DWI to estimate anisotropic and anomalous vascular perfusion, tubular flow, and tissue diffusion in the kidney. Spectral diffusion fit a flexible number of diffusion components, both three-component and two-component diffusion, and returned a flow proxy $fD$ for each distinct component that agreed with clinical biomarkers in renal allografts. Spectral diffusion with a multi-Gaussian model of $fD$ returned component-wise flow proxies that reflected pathological differences in structural tissue, tubules, and vasculature.

The flexibility of spectral diffusion allowed it to detect both two- and three-component simulated anisotropic anomalous diffusion. For three-component diffusion, it performed on a similar or improved level compared to a triexponential and captured tubular flow that a biexponential could not. Further, it was also able to separate two diffusion components on a similar level compared to a biexponential and outperformed the triexponential which returned false components in the presence of noise. However, the Bayesian biexponential was the superior fitting method for the tissue diffusion component. As such, spectral diffusion may be relevant when the number of components cannot be assumed as constant within an organ, and more than just tissue diffusion is of interest. However, if only tissue diffusion is of interest a Bayesian biexponential may be more accurate. Triexponential analysis may benefit from combining exponentials to remove false components.

In simulation, $fD$ showed similar or improved correlation compared to the individual parameters $f$ and $D$. When fitting the signal decay with noise, an overestimated signal fraction $f$ can be countered by an underestimated $D$, and vice versa, meaning combination as $fD$ may adjust for these correlative errors. Agreeing with most current literature, $D$ alone showed limited robustness with the lowest correlation and higher percent error than $f$. In allografts, spectral diffusion $fD$ parameters also detected kidney pathology and showed more significance than individual parameters $f$ and $D$. Spectral diffusion $fD$ as a multi-b-value diffusion parameter may add to the body of knowledge demonstrating diffusion MRI having promise in the evaluation of kidney function and disease[32,37,50].

In allografts, the spectral diffusion parameters agreed with expected cortical pathological changes. $fD_{tubule}$ decreasing with eGFR supports detection of reduced glomerular filtration through reduced flow in kidney tubules. $fD_{vasc}$ decreasing with increased proteinuria supports detection of damaged capillaries or reduced capillary flow that limits reabsorption of protein. $fD_{tubule}$ being significantly lower in impaired allografts supports detection of reduced kidney filtration in tubules[51,52], and $fD_{tissue}$ increasing with IFTA score is consistent with an increased *amount* of slow diffusion due to collagen deposition. While $D_{tissue}$ might be expected to decrease with fibrosis due to greater diffusion restriction from collagen and ECM deposition, the increase in signal fraction $f_{tissue}$ with fibrosis makes the product $fD_{tissue}$ increase with fibrosis. This suggests $fD_{tissue}$ captures the hardening of tissue and accumulation of fibrous tissue by the greater amount of the restricted diffusion, rather than a decrease in the speed of diffusion. While beyond the scope of this work, the sum of all three $fD$ components, or the overall integral of the spectrum, may show a more intuitive reduced total diffusion due to fibrosis. The parallel trends observed between $fD_{tubule}$, and proteinuria, and $fD_{vasc}$ and eGFR, highlight the interconnected nature of dysfunctional allograft physiology and glomerular filtration. While clinically relevant information may be extracted from a translatable multi-b-value DWI sequence (<15 minutes), from $fD$, $f$, and $D$, further work is needed on interpretation and separation of diffusion components.

Regarding comparison to alternative MR perfusion methods, $fD$ from multi-b-value DWI is unique in its methodology being 'local', i.e. independent of blood flow path or delay[53]. Local flow may include all motion, regardless of how it is supplied (e.g. antegrade, delayed, or retrograde) and does not rely on an arterial input function. This has shown benefit in cases of slow and retrograde flow such as strokes[54,55], and allows inclusion of flow other than tagged capillary perfusion such as kidney tubule flow. However, this means direct comparison to standard perfusion MRI (e.g. ASL, DCE, DSC) may be confounded by the fundamentally different methodology[12,14,56,57]. Further, the clinically relevant parameters $fD_{tissue}$ and $fD_{tubule}$ do not have direct corresponding parameters in standard perfusion MRI. Comparison of $fD$ to alternative *in vivo* perfusion MRI measures requires great care and may benefit from modifications to arterial input function selection so both measures are of 'local' perfusion[58,59].

This study is not without limitations. While $fD$ correlated with clinicopathologic values, there may be bias due to kidney volume, kidney weight, hydration status, BMI, sex, race, and age. The simulated kidney cortex needs *in vivo* verification of renal multi-component fractional anisotropy and anomalous diffusion. In addition, the study of kidney medulla components was beyond the scope of this paper and needs further work. This work used b-values≤800s/mm$^2$ for a standard kidney IVIM-DWI sequence, and did not include the effects of more severe diffusion kurtosis which becomes stronger at higher b-values[60]. The simulation did not include the effect of motion correction, T1 or T2 effects, or b-value selection[61]. The spectral diffusion maps presented in this work may be noisier than previous triexponential works because this work presents an image from a single patient scan rather than averaged volunteers[8], and the variable number of components makes zeros indistinguishable from noise. Work is needed on optimal visualization for radiologists such as combining compartments with standardized color maps into a single image[62]. Further, spectral diffusion may not be reliable without high SNR from averaging, denoising, and a 3T magnet. The simulated anomalous diffusion was deduced from values in the brain, assumed ballistic perfusion, and did not include the effects of different types of anomalous diffusion[18,20,63]. Multi-component anomalous diffusion interpretation may benefit from more complex, comprehensive modeling, Monte Carlo simulations, and *in vivo* imaging [28,48,63]. While this work demonstrated agreement with kidney pathology at the clinical subgroup level, validation is needed from a larger clinical study. Clinical application may benefit from more robust measurement of the non-Gaussian diffusion with T2-correction[64,65], varying gradient durations, flow-compensation, directions, and higher b-values such as multi-shell diffusion tractography[60,66-68].

A data-driven approach of spectral diffusion demonstrated a clinically translatable multi-b-value DWI sequence in simulation and clinical application of kidney cortex. Simulation of multi-component anisotropic and anomalous diffusion demonstrated that $fD$ as a flow proxy returned fair correlation to input truth. Application in kidney allografts showed $fD$ agreeing with kidney allograft pathophysiology, including eGFR, proteinuria, and fibrosis scores, supporting potential clinical utility of $fD$ as a flow proxy from multi-b-value spectral diffusion.

**Data and Code Availability:** Data is available upon reasonable request to the corresponding author. Open-source code developed to generate spectral volumes and multi-component flow maps described in this paper is publicly available on github along with a de-identified example data set: https://github.com/miramliu/Spectral_Diffusion.

## Tables and Figure Captions

Table 1. Simulation of multi-component diffusion with biexponential, triexponential, and spectral diffusion. Presented are mean and standard deviation of the normal distributions that the input parameters of the simulation were chosen from. Component fractions were normalized to ensure the components summed to one.

| **Distributions of simulated anisotropic Gaussians** | **Tissue component** | **Tubule component** | **Vascular component** |
|---|---|---|---|
| Signal fraction (*f*) | 0.60± 0.10 | 0.30±0.015 | 0.10±0.05 |
| Diffusion coefficient ($D_{trace}$) | 0.0015±0.00075 | 0.010±0.0025 | 0.070±0.009 |
| Fractional Anisotropy (FA) | 0.18±0.02 | 0.12±0.03 | 0.09±0.04 |
| Anomalous $\gamma$ | 0.85±0.051 | 1.0±0.10 | 1.75±0.2 |

Table 2. The three and two-component anisotropic anomalous diffusion simulation comparing individual signal fraction and diffusion coefficient from the three fitting methods. Presented is the linear regression between input parameters and the output fit parameters, and the median percent difference (Δ%) of each component. Note that for tubule component of a three-component model, the biexponential automatically returns an incorrect $f_{tubule}$ and $D_{tubule}$ of zero (see 200% error in three-component biexponential). Similarly, in the two-component model, the triexponential returns an incorrect non-zero $f$ and $D$.

| | **Biexponential** | **Triexponential** | **Spectral** |
|---|---|---|---|
| Three-component $f$ | $y = 0.65x + 0.1, R^2 = 0.26$ $[\Delta f_{vasc}, \Delta f_{tubule}, \Delta f_{tissue}] = [112.0\%, 200\%, 5.9\%]$ | $y = 0.39 + 0.2, R^2 = 0.17$ $[\Delta\Delta f_{vasc}, \Delta f_{tubule}, \Delta f_{tissue}] = [56.9\%, 39.5\%, 29.3\%]$ | $y = 0.44x + 0.2, R^2 = 0.22$ $[\Delta\Delta f_{vasc}, \Delta f_{tubule}, \Delta f_{tissue}] = [39.5\%, 30.3\%, 29.9\%]$ |
| Three-component $D$ | $y = 0.30x - 0.7, R^2 = 0.79$ $[\Delta D_{vasc}, \Delta D_{tubule}, \Delta D_{tissue}] = [111.0\%, 200.0\%, 13.6\%]$ | $y = 0.72x - 0.8, R^2 = 0.24$ $[\Delta D_{vasc}, \Delta D_{tubule}, \Delta D_{tissue}] = [66.3\%, 53.2\%, 58.9\%]$ | $y = 1.05x - 1.9, R^2 = 0.55$ $[\Delta D_{vasc}, \Delta D_{tubule}, \Delta D_{tissue}] = [22.0\%, 37.8\%, 40.8\%]$ |
| Two-component $f$ | $y = 0.82x + 0.1, R^2 = 0.98$ $[\Delta f_{vasc}, \Delta f_{tissue}] = [33.0\%, 7.2\%]$ | $y = 0.15x + 0.3, R\hat{}2 = 0.04$ $[\Delta f_{vasc}, \Delta f_{tissue}] = [18.1\%, 75.9\%]$ | $y = 0.90x + 0.0, R\hat{}2 = 0.89$ $[\Delta f_{vasc}, \Delta f_{tissue}] = [17.2\%, 4.0\%]$ |
| Two-component $D$ | $y = 0.65x + 0.5, R^2 = 0.81$ $[\Delta D_{vasc}, \Delta D_{tissue}] = [40.2\%, 16.1\%]$ | $y = 0.72x + 5.1, R^2 = 0.41$ $[\Delta D_{vasc}, \Delta D_{tissue}] = [34.2\%, 66.1\%]$ | $y = 1.02x - 0.6, R^2 = 0.63$ $[\Delta D_{vasc}, \Delta D_{tissue}] = [21.1\%, 13.3\%]$ |

Table 3. The three and two-component anisotropic anomalous diffusion simulation correlating $fD$ of output parameters from the three fitting methods to $fD$ of input parameters. Presented is the linear regression between input parameters and the output fit parameters. Note that for tubule component of a three-component model, the biexponential automatically returns an incorrect $f_{tubule}$ and $D_{tubule}$ of zero (see 200% error in three-component biexponential). Similarly, in the two-component model, the triexponential returns an incorrect non-zero $f$ and $D$.

Bland-Altman mean difference and 95% confidence interval BA(95%CI) between the input $fD$ and the output fit $fD$ parameters is included in units of $10^{-3}$ mm$^2$/s.

| | **Biexponential** | **Triexponential** | **Spectral Diffusion** |
|---|---|---|---|
| Three-component $fD$ | $y = 0.93x - 0.6, R^2 = 0.67$<br>$BA(95\%CI) =$<br>$-0.85(-4.9, 3.2)$<br>$[\Delta fD_{vasc}, \Delta fD_{tubule}, \Delta fD_{tissue}]$ $= [22.8\%, 200\%, 20.6\%]$ | $y = 0.81x - 0.0, R^2 = 0.73$<br>$BA(95\%CI) =$<br>$-0.74(-3.8, 2.4)$<br>$[\Delta fD_{vasc}, \Delta fD_{tubule}, \Delta fD_{tissue}]$ $= [27.3\%, 31.2\%, 90.2\%]$ | $y = 1.10x - 0.1, R^2 = 0.74$<br>$BA(95\%CI) =$<br>$0.23(-3.8, 4.3)$<br>$[\Delta fD_{vasc}, \Delta fD_{tubule}, \Delta fD_{tissue}]$ $= [20.0\%, 28.3\%, 76.3\%]$ |
| Two-component $fD$ | $y = 0.95x - 0.1, R^2 = 0.93$<br>$BA(95\%CI) =$<br>$-0.39(-3.3, 2.5)$<br>$[\Delta fD_{vasc}, \Delta fD_{tissue}]$ $= [23.1\%, 14.9\%]$ | $y = 0.63x + 0.7, R^2 = 0.65$<br>$BA(95\%CI) =$<br>$-0.84(-9.2, 7.5)$<br>$[\Delta fD_{vasc}, \Delta fD_{tissue}]$ $= [31.3\%, 167.7\%]$ | $y = 1.01x + 0.2, R^2 = 0.88$<br>$BA(95\%CI) =$<br>$0.29(-3.9, 4.5)$<br>$[\Delta fD_{vasc}, \Delta fD_{tissue}]$ $= [15.8\%, 14.9\%]$ |

Table 4. Average values and standard deviations of the individual parameters $f$ and $D$ for spectral diffusion, biexponential, and triexponential fits across IFTA score (range 0-6) for allografts presenting with normal function (eGFR≥45 ml/min/1.73m$^2$) for all components. One-way ANOVA of each parameter across fibrosis scores is provided.

| **Parameter** | **IFTA=0** | **IFTA=2** | **IFTA=4** | **IFTA=6** | **F-stat, p-value** |
|---|---|---|---|---|---|
| **Spectral Diffusion** | | | | | |
| $f_{vasc}$ | 0.07±0.03 | 0.05±0.04 | 0.02±0.04 | 0.006±0.05 | F-stat=1.06, p=0.38 |
| $f_{tubule}$ | 0.09±0.04 | 0.06±0.04 | 0.02±0.07 | 0.04±0.02 | F-stat=2.19, p=0.11 |
| $f_{tissue}$ | 0.75±0.07 | 0.83±0.08 | 0.89±0.11 | 0.85±0.08 | F-stat=2.89, p=0.06 |
| $D_{vasc}$ | 87.7±29.20 | 75.6±43.42 | 42.3±49.15 | 12.85±65.74 | F-stat=1.67, p=0.20 |
| $D_{tubule}$ | 7.10±2.66 | 4.17±6.19 | 1.32±2.92 | 2.18±2.15 | F-stat=2.39, p=0.09 |
| $D_{tissue}$ | 2.02±0.27 | 2.14±0.19 | 1.98±0.27 | 2.69±0.49 | F-stat=2.89, p=0.06 |
| **Biexponential** | | | | | |
| $f$ | 0.21±0.03 | 0.19±0.03 | 0.19±0.02 | 0.24±0.03 | F-stat=2.49, p=0.08 |
| $D$ | 1.54±0.26 | 1.57±0.17 | 1.68±0.17 | 1.86±0.15 | F-stat=1.07, p=0.38 |
| $D^*$ | 37.00±14.00 | 31.96±10.87 | 19.56±13.39 | 19.08±14.91 | F-stat=0.97, p=0.42 |
| **Triexponential** | | | | | |
| $f_{vasc}$ | 0.09±0.03 | 0.06±0.04 | 0.03±0.04 | 0.07±0.03 | F-stat=1.48, p=0.24 |
| $f_{tubule}$ | 0.46±0.09 | 0.44±0.06 | 0.54±0.12 | 0.46±0.15 | F-stat=0.35, p=0.78 |
| $f_{tissue}$ | 0.47±0.08 | 0.53±0.08 | 0.60±0.14 | 0.57±0.19 | F-stat=0.53, p=0.66 |
| $D_{vasc}$ | 182.95±38.26 | 153.20±59.64 | 96.72±46.91 | 184.4±65.26 | F-stat=2.62, p=0.07 |
| $D_{tubule}$ | 5.04±1.29 | 3.85±1.93 | 2.59±1.94 | 4.52±3.78 | F-stat=2.14, p=0.12 |
| $D_{tissue}$ | 1.14±0.34 | 1.23±0.21 | 1.33±0.32 | 1.62±0.57 | F-stat=0.81, p=0.51 |

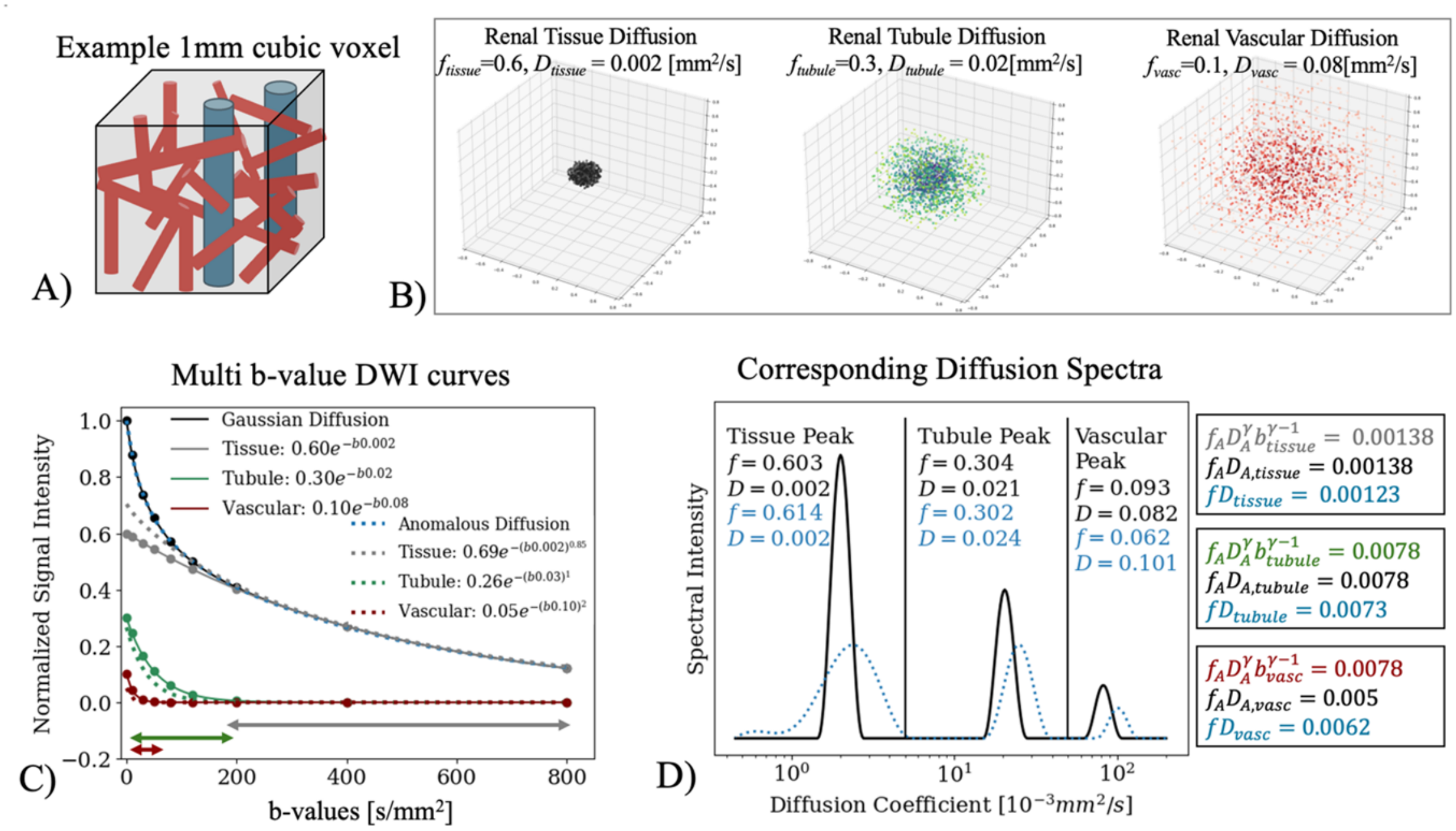

Figure 1. (A) Example components of a renal voxel with tissue (grey), tubules (green), and vascular (red) components. (B) Simulated isotropic Gaussian diffusion over 100 seconds from the center of a cubic mm. (C) The DWI multi- b-value curve as a triexponential of three Gaussian components (solid: black) along with corresponding individual components (solid: grey, green, red) with the legend showing each $fe^{-bD}$ with $f, D$ as the numeric values in (B). Anomalous diffusion (dashed: blue) is plotted as the sum of three anomalous diffusion components as stretched exponents (dashed: grey, green, red) that would return a similar multi b-value decay curve as a multi-Gaussian. The corresponding individual anomalous diffusion components curves are plotted (dashed: grey, green, red). Colored lines along the bottom demarcate the range of b-values that in which the three simulated diffusion components are dominant (tissue:grey, tubular:green, vascular:red,). (D) Shows the spectrum of the simulated multi b-value curve and the fractions and diffusion coefficients from the spectrum (black: Gaussian; blue: anomalous) of the three returned peaks. Vertical lines separate the three diffusion regimes of this work (0.8 < tissue < 5 ≤ tubule < 50 ≤ vascular in $10^{-3}$ mm²/s). Following Eq. 8, numeric values of $f_A D_A^{\gamma} b^{\gamma-1}$ (grey) with the b-value selected from the center of each diffusion regime shown in Fig.1C (grey: $b_{tissue} = 500$, green:$b_{tubule}$=100, red:$b_{vasc} = 10$), as well as $f_A D_A$ (black), and corresponding spectral diffusion $fD$ (blue) from this example are shown for the three diffusion components from the three different diffusion regimes in text boxes on the right.

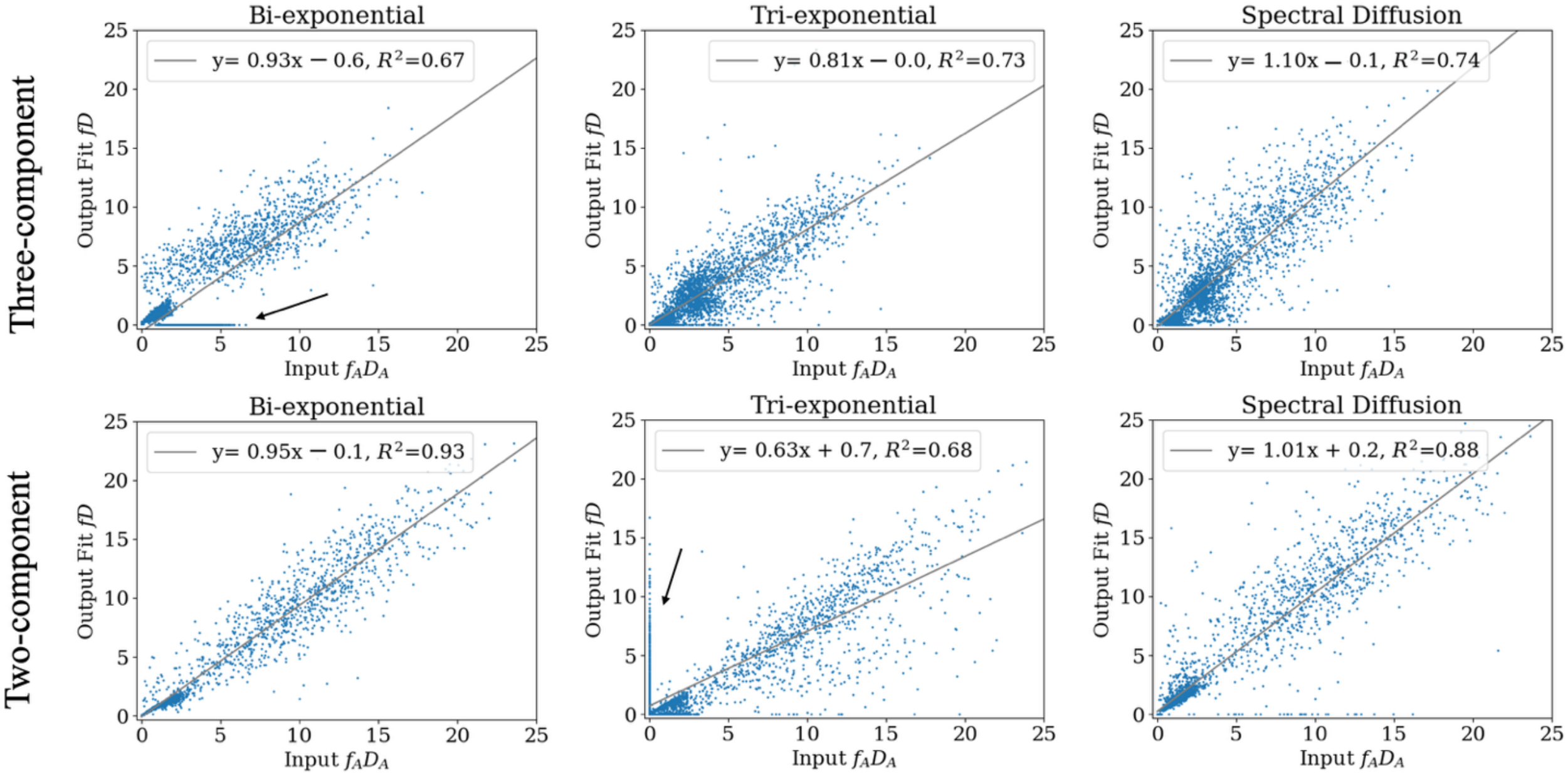

Figure 2. Linear regression correlation plots for anisotropic anomalous diffusion for three-component (top row) and two-component (bottom row) simulations. Units are signal fraction $f$ times $D$ in $10^{-3}$ $mm^2/s$. Black arrows note the false zero $fD$s from the biexponential fit of a three-component simulation (upper left), and the false non-zero $fD$s from the triexponential fit of a two-component simulation (bottom center).

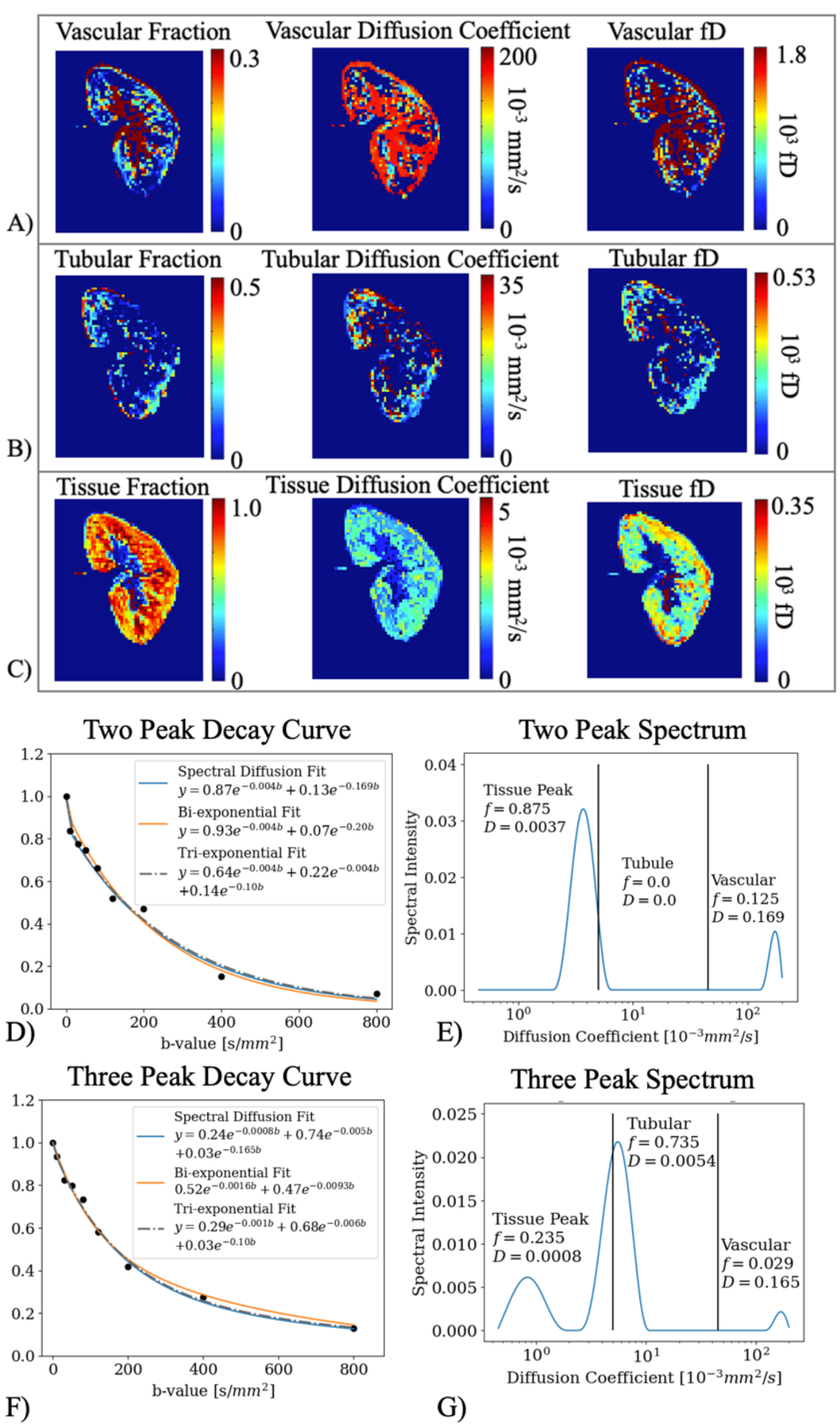

Figure 3. Images from a volunteer native kidney (M/47y) for (A) vascular spectral diffusion parameter and flow maps, (B) tubular spectral diffusion parameter and flow maps, and (C) tissue parenchyma spectral diffusion parameter and flow maps. Note the difference in scale for each component and parameter. (D) A DWI decay curve overlaid with the spectral diffusion fit, biexponential fit, and triexponential fit of a volunteer cortical voxel that returned two spectral peaks. (E) The corresponding spectrum of (D) labeled with the fraction and diffusion coefficient, which were used to plot the spectrum in subfigure (D); vertical lines demonstrate the assigned divisions into different physiologic components. (F-G) is the same as (D-E) for a volunteer cortical voxel that returned three spectral peaks. Note the difference in scales on the y axis between E and G.

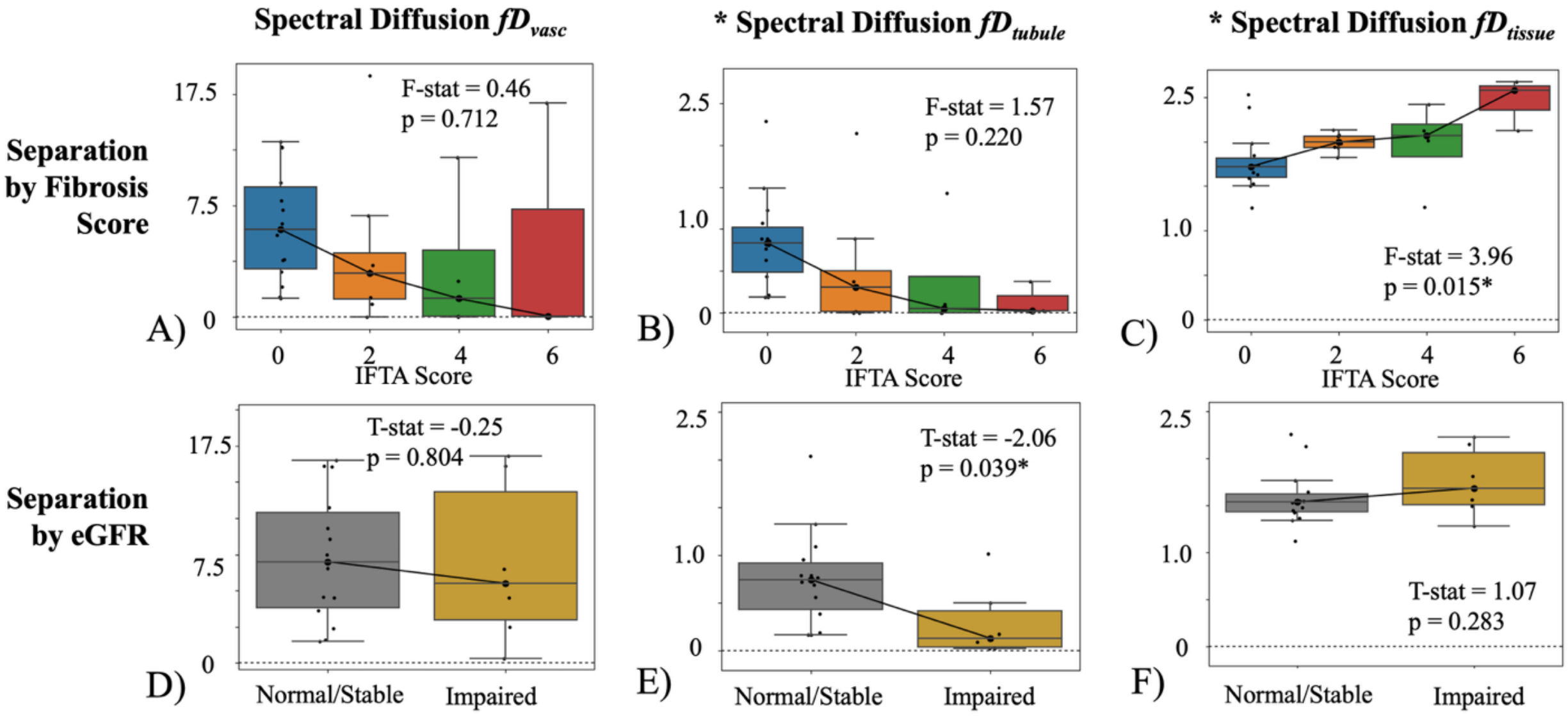

Figure 4. Spectral diffusion $fD$ across IFTA score (range 0-6) for allografts presenting with normal function (eGFR$\geq$45 ml/min/1.73m$^2$) with (A) vascular component, (B) tubular component, and (C) tissue parenchyma component. Multi-component $fD$ of impaired-function allografts (eGFR<45 ml/min/1.73m$^2$) and normal/stable allografts (eGFR$\geq$45 ml/min/1.73m$^2$) for allografts with no fibrosis (IFTA=0) with (D) vascular component (E) tubular component, and (F) the tissue component. Note the different scales across the different components. An asterisk marks those that are statistically significant (p<0.05). Individual data points are plotted and the number of patients per each subgroup is included in Supporting Information S2, Table 1, e.g. 14 patients with IFTA = 0 and healthy function for the blue box in subfigure A. The number of patients still presenting with normal function decreases at higher IFTA scores, as expected.

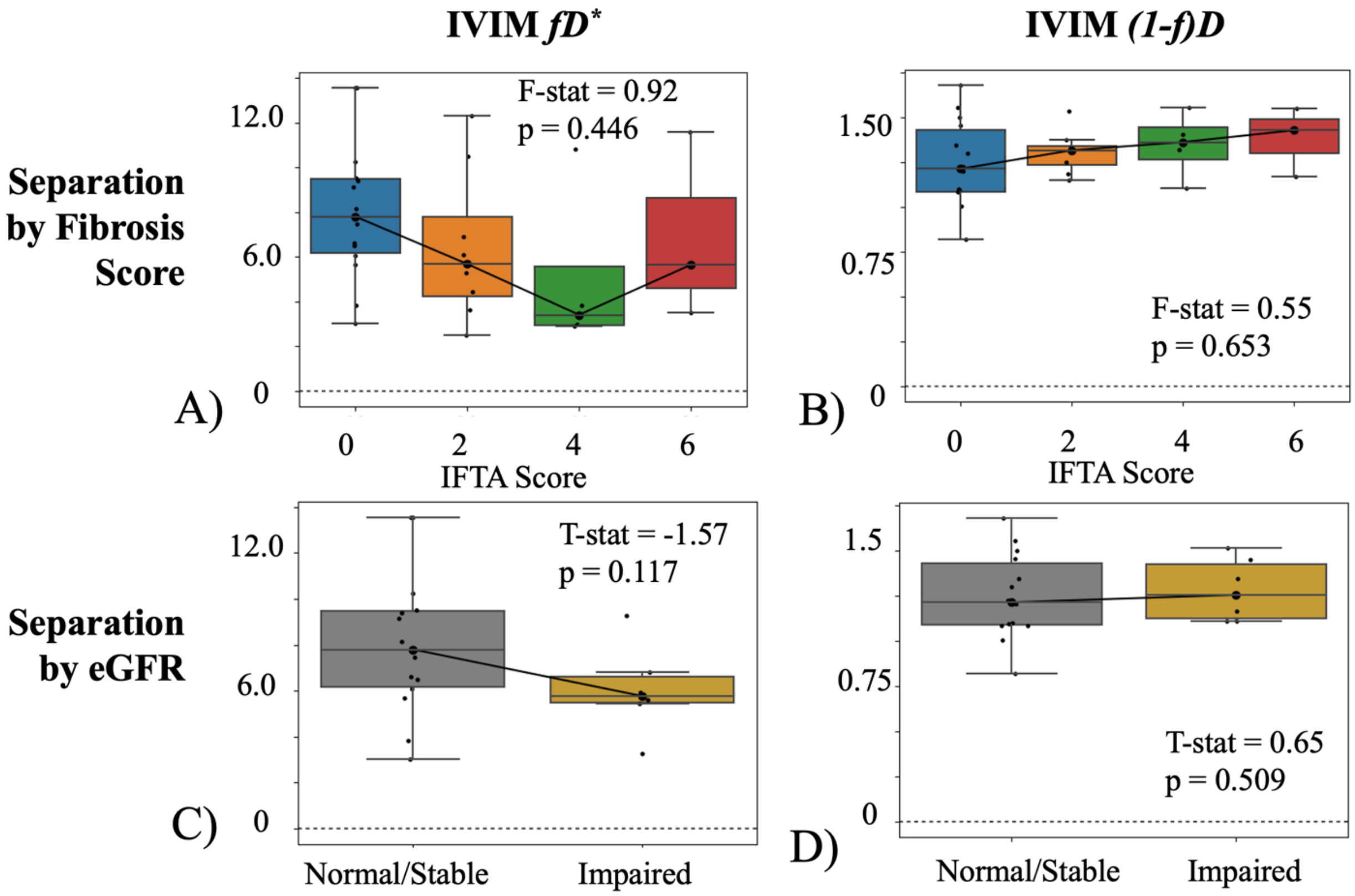

Figure 5. Biexponential IVIM across IFTA score (range, 0-6) for allografts presenting with normal/stable function (eGFR≥45 ml/min/1.73m$^2$) with (A) $fD_{vasc} \equiv fD^*$ and (B) tissue parenchyma $fD_{tissue} \equiv (1\text{-}f)D$. Multi-component $fD$ of impaired allografts (eGFR<45 ml/min/1.73m$^2$) and normal/stable allografts (eGFR≥45 ml/min/1.73m$^2$) for allografts with no fibrosis (IFTA=0) with (C) vascular component and (D) the tissue component. Note the different scales across the different components. Individual data points are plotted and the number of patients per each subgroup is included in Supporting Information S2,, e.g. 14 patients with IFTA = 0 and healthy function for the blue box in subfigure A. The number of patients still presenting with normal function decreases at higher IFTA scores, as expected.

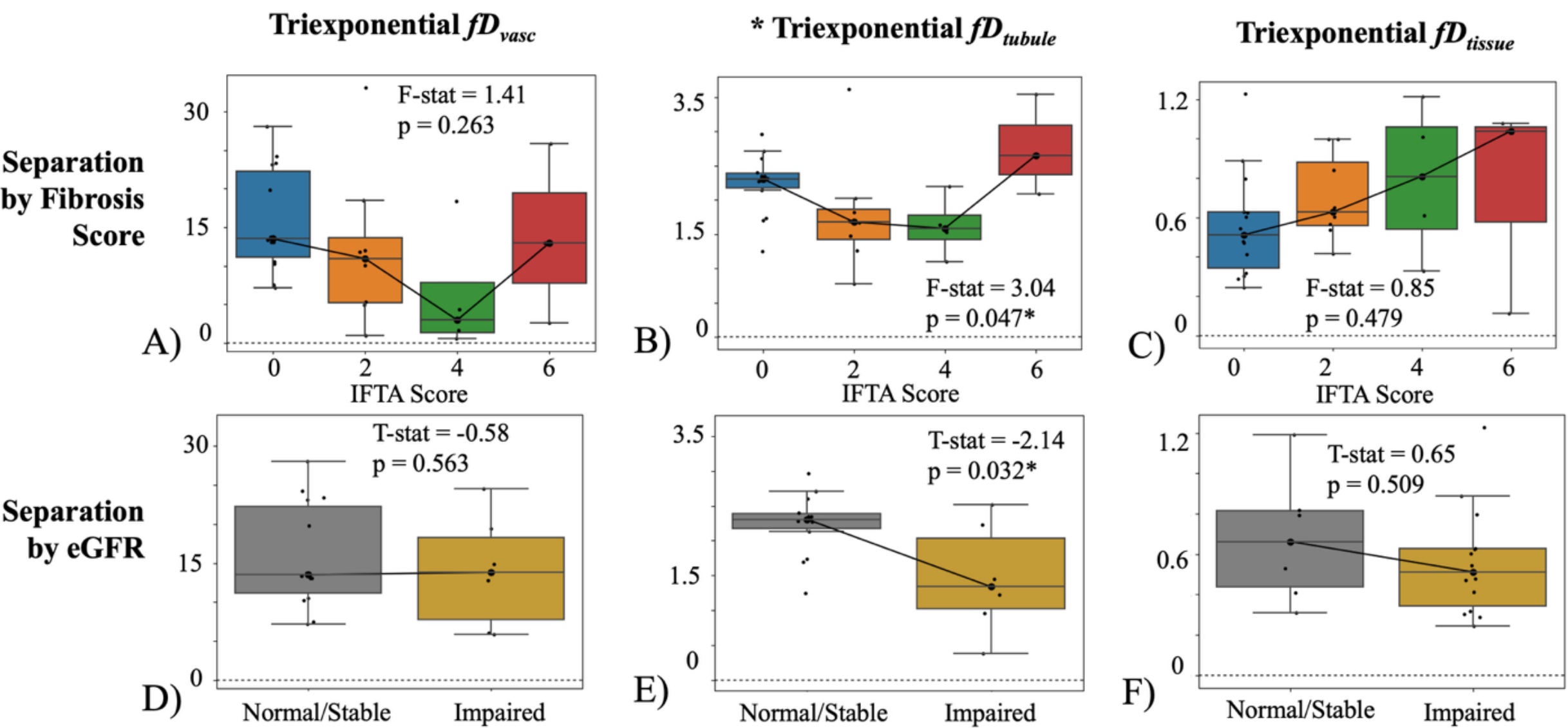

Figure 6. Triexponential *fD* across IFTA score (range, 0-6) for allografts with normal/stable function (eGFR≥45 ml/min/1.73m$^2$) with (A) vascular component, (B) tubular component, and (C) tissue parenchyma component. Multi-component *fD* of impaired allografts (eGFR<45 ml/min/1.73m$^2$) and normal/stable allografts (eGFR≥45 ml/min/1.73m$^2$) for allografts without fibrosis (IFTA=0) with (D) vascular component (E) tubular component, and (F) the tissue component. Note the different scales across the different components. An asterisk marks those that returned statistically significant difference at the p = 0.05 level. Individual data points are plotted and the number of patients per each subgroup is included in Supporting Information S2, Table 1, e.g. 14 patients with IFTA = 0 and healthy function for the blue box in subfigure A. The number of patients still presenting with normal function decreases at higher IFTA scores, as expected.

**Supporting Information S1.** Influence of Spectral Diffusion Regularization Factor λ Selection
The weighting factor λ can be selected by generalized cross-validation[69,70] for every decay curve, or fixed as a constant for all analyses if signal-to-noise-ratio (SNR) is stable. A lower λ allows for sharper spectral peaks while higher λ returns broader spectra for lower SNR. Spectral diffusion was run with the regularization parameter fixed at $\lambda_{0.1} = 0.1, \lambda_2 = 2, \lambda_8 = 8$, and $\lambda_{CV}$ = cross-validated from optimal $\lambda \approx \frac{\#bval}{SNR}$[13]. Multi-component fD using $\lambda_{CV}$, $\lambda_{0.1}$, or $\lambda_8$ showed no significant difference in correlation or linear regression to truth in simulation. All simulations were run on a 12 Core Apple M2 Max running MATLAB R2025a (MATLAB Engine API, 2025) and Python 3.11 (Anaconda Inc); fixed λ reduced the computation time 7.4-fold for $\lambda_{0.1}$ and 4.8-fold for $\lambda_8$ for a three-component model (9.8-fold and 4.5-fold for two-component, respectively). For fixed regularization of the spectral map in Fig.3, lower λ correlated linearly with reduced computation times ($t_\lambda(s) = 11.07\lambda + 27.3, p < 0.0001, R^2 = 0.95$). Compared to $\lambda_{CV}$ fixed-regularization had a 208-fold and 29-fold decrease in computation time ($t_{\lambda_{CV}}$ =3334s, $t_{\lambda_8} = 115s$, $t_{\lambda_{0.1}} = 16s$).

For the renal allograft MRIs, with b0 SNR=50, voxels most commonly returned two-peak and three-peak spectra, with the following percentages for $\lambda_{0.1}$, $\lambda_8$ and $\lambda_{CV}$ respectively: 1-peak=[14.8%,18.7%,18.6%], 2-peak=[37.3%,42.2%,42.4%], 3-peak=[32.3%,34.8%,34.8%,], 4-peak=[4.1%,2.4%,2.3%]. These were calculated as the number of peaks (i.e. signal with decreasing values on either side after smoothing, see Fig 1D. for example) in each voxel's spectrum. After sorting the peaks, multi-component flow agreed between the three λs (CCC = 0.99, 1.00 for fD from fits with $\lambda_{0.1}, \lambda_8$ against $\lambda_{CV}$ respectively supporting use of $\lambda_{0.1}$ for faster computation. Visually, parameter maps did not vary noticeably between different regularization factors. Therefore, in this work, use of a fixed regularization term with $\lambda_{0.1}$ ensured consistency in fitting method across voxels, scans, and patients. As more smoothing may be needed for signal with lower SNR, $\lambda > 0.1$ may be better for processing sequences with SNR<50.

**Supporting Information S2.** Clinical Demographics
Table 1. Patient demographics. Patients included in the study were those enrolled from 2/2022-09/2024 who are >1-month post-transplant. Informed consent was obtained, and patients underwent a non-contrast MRI protocol within 7 days of biopsy that included advanced DWI. Exclusion criteria were age <18 years, large vessel or urinary tract complication of the kidney transplant, contra-indications to MRI, or pre-existing medical conditions including a likelihood of developing seizures or claustrophobic reactions.

| **Demographics and Clinical Features** | |
|---|---|
| Biopsy Type | |
| Indication Biopsy | N=45 |
| Protocol Biopsy | N=9 |
| Sex(F/M) | 21/33 |
| Race | |
| Black/African American | 30 |
| White | 5 |
| Asian | 6 |
| Other/Unchecked | 13 |
| Age (years; mean±std, range) | 48.8±10.5 (25-66) |
| Weight (kg; mean±std, range) | 78.61±16.84 (44.5-118.8) |
| BMI ($kg/m^2$, mean±std, range) | 27.6±5.1 (17.4-38.2) |
| Allograft Volume (mL, mean±std, range) | 241±78 (100-573) |
| Living Donor | N=20; |
| Donor Age (years; mean±std, range) | 36.41±11.33 (21-63) |

| Deceased Donor | N=34; |
|---|---|
| KDPI | All KDPI < 85 |
| Time since transplant (months; mean±std, range) | 47.2±66.1 (1.3-252) |
| U-Protein (mg/24hr; mean±std, range) | 159.82±284.63 (1.0-1513), 8 unknown |
| Donor Specific Antibodies status | 42 negative, 6 positive, 6 unknown |
| **eGFR (CKD-EPI 2021 mL/min/1.73$m^2$; mean±std, range)** | 44.96±19.30 (8.0-105.0) |
| eGFR< 45 (mL/min/1.73$m^2$; mean±std, range) | N=26; 28.67±9.96 (8.0-44.0) |
| eGFR≥ 45 (mL/min/1.73$m^2$; mean±std, range) | N=28; 59.01±13.49 (45.0-105.0) |
| **Interstitial Fibrosis/ Tubular Atrophy (IFTA)** | Total number/number with healthy eGFR |
| IFTA = 0 | 20/14 |
| IFTA = 2 | 13/8 |
| IFTA = 4 | 10/4 |
| IFTA = 6 | 11/3 |
| **Clinical subgroups** | |
| Normal/stable function and no fibrosis: eGFR≥ 45 & IFTA=0 | 14 |
| Normal/stable function and fibrosis: eGFR≥ 45 & IFTA>0 | 15 |
| Impaired function and no fibrosis: eGFR< 45 & IFTA=0 | 6 |
| Impaired function and fibrosis: eGFR< 45 & IFTA>0 | 19 |